\documentclass[manuscript]{aastex}

\received{}
\revised{}
\accepted{}

\shortauthors{Jones {\it et al.}}
\shorttitle{SW 3}

\begin{document}

\title{Evidence of Fragmenting Dust Particles from Near-Simultaneous Optical and Near-IR Photometry and Polarimetry of Comet 73P/Schwassmann-Wachmann 3}
\author{Terry Jay Jones~\altaffilmark{1}, David Stark, Charles E. Woodward}
\affil{Department of Astronomy, University of Minnesota \\ 116 Church 
Street S.E., Minneapolis, MN 55455 \\ tjj@astro.umn.edu, star0288@umn.edu, chelsea@astro.umn.edu}
\author{Michael S. Kelley}
\affil{Department of Physics, University of Central Florida \\ 4000 Central Florida Boulevard,
Orlando, FL 32816-2385 \\ msk@physics.ucf.edu}
\author{Ludmilla Kolokolova}
\affil{Department of Astronomy, University of Maryland \\ College Park, MD, ludmilla@astro.umd.edu}\author{Dan Clemens, April Pinnick}
\affil{Institute for Astrophysical Research, Boston University \\ 725 Commonweath Ave., Boston, MA, clemens@bu.edu, apinnick@bu.edu}

\altaffiltext{1}{Visiting Astronomer at the Infrared Telescope Facility which is operated by the University of Hawaii under contract from the National Aeronautics and Space Administration.}

\begin{abstract}

We report imaging polarimetry of segments B and C of the Jupiter-family Comet 73P/Schwassmann-Wachmann 3 in the $I$ and $H$ bandpasses at solar phase angles of approximately 35 and $85\degr$. The level of polarization was typical for active comets, but larger than expected for a Jupiter-family comet. The polarimetric color was slightly red $(\frac{\delta P}{\delta \lambda}=+1.2\pm 0.4)$ at a phase angle of $\sim 35 \degr$ and either neutral or slightly blue at a phase angle of $\sim 85 \degr$. Observations during the closest approach from 2006 May 11-13 achieved a resolution of $35$~km at the nucleus. Both segments clearly depart from a $1/\rho$ surface brightness for the first $50-200$~km from the nucleus. Simulations of radiation driven dust dynamics can reproduce some of the observed coma morphology, but only with a wide distribution of initial dust velocities (at least a factor of 10) for a given grain radius. Grain aggregate breakup and fragmentation are able to reproduce the observed profile perpendicular to the Sun-Comet axis, but fit the observations less well along this axis (into the tail). The required fragmentation is significant, with a reduction in the mean grain aggregate size by about a factor of 10. A combination of the two processes could possibly explain the surface brightness profile of the comet. 

\end{abstract}

\keywords{comets: polarimetry --- comets: individual (73P)} 

\section{Introduction} 

The close apparition of comet 73P/Schwassmann-Wachmann 3 (hereafter SW3) was a highly anticipated astronomical event of 2006. First, the comet was unusually close: on 2006 May 12, it passed the Earth at a distance of 0.08 AU ($1.2 \times 10^7$~km). Second, SW3 is among the most active comets of the currently known Jupiter-family comets. In all other ways, this comet is a typical Jupiter-family comet with aphelion and perihelion distances equal correspondingly to 5.187 AU, and 0.9391 AU,  orbital period T = 5.36 years, and orbit inclination of 11.39\degr.

Observations in 1995 (two orbital periods before the 2006 apparition) showed that upon reaching perihelion SW3 suddenly increased in brightness \citep{cro96} and fragmented producing three bright and many fainter components \citep{sco96}. During the 2006 apparition, fragmentation continued and over 60 fragments of the comet were detected \citep{wea06}. The intrinsic brightness of many of these fragments changed with time, indicating that fragmentation continued on smaller scales. The fragmentation of the comet nucleus exposed the interior of the progenitor, thus, truly pristine materials from the comet interior could be observed \citep{rus07}. We consider this apparition of SW3 as a natural analog of the artificial experiment performed by the Deep Impact mission \citep{ahe05}, which provided us with opportunity to observe material released from the interior of a the nucleus of 9P/Tempel \citep{har05, har07, mee05}. \Citet{fur07} made imaging polarization observations of dust expelled by 9P/Tempel after impact that showed a high polarization suggestive of small grains. The dust observed in the SW3 fragments might be expected to be different from the dust regularly observed from Jupiter-family comets, which likely represents larger grains of highly processed material arising from the nucleus surface layers.

Polarimetry is a powerful method for studying properties of cometary dust. Polarimetry provides information about the dust size distribution, structure, shape, and composition of the dust particles \citep[c.f., reviews by][]{joc97, kis99, had03a, kol04a}. \citet{kis02} and \citet{had03b} show that the polarization of comets can change dramatically during break-up, indicating exposure of materials with physical characteristics different from the interior of the nucleus. The apparition of comet SW3 in 2006 presented an opportunity to obtain unique data on properties of dust produced by a comet undergoing fragmentation. Conclusions based on the polarimetric measurements can be significantly improved if simultaneous photometric images of the coma are also available. The surface brightness images allow us to see jets, shells and other coma features, and then to connect them with the features in polarimetric images. 

The angular dependence of polarization (polarization vs. phase angle) is used to extract physical properties of the dust from the polarimetric data. For instance, \citet{gus99} demonstrate that the spectral dependence of polarization can significantly increase our understanding of cometary dust properties. The vast majority of the comet polarimetric data have been obtained in the visible wavelengths, and a definitive behavior in the near-infrared has not been established. The spectral gradient of polarization (polarimetric color) is a powerful diagnostic tool for revealing the size and structure of cometary grains \citep{kol04a}. A positive (red) value in the visible, typical for the majority of comets, is strong evidence for the presence of loose aggregates in cometary dust \citep{kol04b}. However, observations in the near infrared suggest a trend to negative values at longer wavelengths \citep{kel04, jg, has97}, at least in some comets. A study of the detailed behavior of the comet polarimetric color over a wide wavelength range from visible to near-infrared provides light-scattering modelers with important information to further constrain properties of cometary grains.

Simulation of dust ejection and subsequent flight paths are able to reproduce some of the morphology of comet comae \citep[e.g.,][]{kel06, ful04}. Fragmentation of comet nuclei have also been studied and large scale fragmentation can clearly be seen in images of comets such as D/1993 F2 (Shoemaker-Levy 9) \citep{wea95}. However, fragmentation of small grain aggregates is difficult to ascertain from images with $1000$~km scales. HST images of SW3 segment B show dramatic fragmentation of the nucleus \citep{wea06}. 

The close approach of SW3 and our ability to make nearly simultaneous visual and near-infrared polarimetric observations motivated us to make the observations reported in this paper. We seek to explore evolution of the scattering properties of cometary dust at physical size scales smaller than possible with most other comets. Our ability to make polarimetric images at two widely spaced wavelengths allows us to explore the wavelength dependence of the polarization without resorting to comparisons between observations at different epochs through different bandpasses. We find that our observations of SW3 at $50$~km scales can not be explained by simple dust ejection alone, but show evidence for grain fragmentation within a few minutes of release from the nucleus.

\section{Observations} 

Observations were conducted using three instruments on three different telescopes. An observing log, including observation time, telescope, and instrument is found in Table 1. The infrared observations on 2006 May 12-13 UT were made with NSFCAM2 in polarimetry mode on the NASA IRTF 3-M telescope at a plate scale of $0\farcs04$ per pixel, resulting in a field of view of $80\arcsec$ square. In polarimetry mode, NSFCAM utilizes a rotating half-wave plate at the entrance window of the camera and a cold wire grid polarizer in the second filter wheel. Details of the NSFCAM (+Polarimeter) observing technique are given in \citet{jon00} and \citet{kel04}.

The infrared observations on 2006 April 18 and 19 UT were made using Mimir, a cryogenic, facility-class instrument for conducting wide-field imaging, long-slit spectroscopy, and imaging polarimetry on the Perkins 1.83 m telescope outside Flagstaff, Arizona \citep{clem07}. We used the f/5 camera, resulting in a plate scale of $0\farcs58$ per pixel. Polarimetry mode utilizes a cold, rotating zero-order half-wave plate in $H$ $(1.65~\micron$ band and a fixed wire grid for analysis. The detector is a 1024x1024 InSb Aladdin III hybrid device with 32 parallel readout channels. Our data reduction followed the same procedure as with NSFCAM2.

All of the optical polarimetry was performed using OPTIPOL, an optical polarimeter at the University of Minnesota Mount Lemmon Observing Facility. The polarimeter utilizes a $1024\times1024$ CCD (Santa Barbara Instrument Group ST-1001E) with a plate scale of $0\farcs25$ per pixel. A half-wave plate rotates the polarization of the incoming light and a Wollaston prism splits the incoming image into two images with perpendicular polarizations. Both images are recorded simultaneously to reduce errors caused by fluctuations in the atmosphere. All optical polarimetry was done using a Cousins-Kron $I$ band filter. Normally we would have used a narrow-band $\lambda_0 = 0.676\pm0.01~\micron$ filter, but SW 3 was too faint. 

\section{Data Reduction}

Methods for reducing comet imaging polarimetry data are discussed in \citet{kel04} and \citet{jg}. Since the two primary components of SW3 (B and C) were relatively small compared to our total field of view, we chose the method that forces the sky well away from the comet to have $Q=U=0$. OPTIPOL uses a Wollaston prism to split the beam into two orthogonal polarizations, each imaged on the CCD. By swapping the two polarizations using a rotating half-waveplate, differential effects can be removed. We find no evidence for any residual instrumental polarization above the $0.1\%$ level due to the telescope and camera optics. Our previous work using the original NSFCAM on the IRTF showed no measurable instrumental polarization to the $0.1\%$ level across the entire field of view. With the NSFCAM2 upgrade, some of the optics have changed and this has led to a modest instrumental polarization. 

To measure the instrumental polarization for NSFCAM2 we observed the globular cluster M13, which fills the field of view with stars that are likely to be either unpolarized or weakly polarized. Based on ten stars in M13, the mean instrumental polarization was $3.2\%$ with variations of $\pm 0.3\%$ across the field. We did not have the time to pursue a more accurate determination of the spatial variations in the instrumental polarization, so we are limited in accuracy to uncertainties in these variations at the $\pm 0.3\%$ level. All of our NSFCAM2 polarimetry will carry this basic systematic error. Both Mimir and NSFCAM2 observations used the photometric standard HD 136753 \citep{eli82}.

For Mimir, we used the instrumental polarization calibration available from the instrument web site. The comet was observed by taking images at different locations on the Mimir detector, and the instrumental polarization specific to those locations was used to correct the raw data. At all of these locations in the focal plane the corresponding instrumental polarization was less than $0.5\%$. There are variations in the position angle of the instrumental polarization across the field of view in Mimir \citep{clem07}. This was taken into account when subtracting the instrumental polarization. Since SW3 was relatively faint at the time of our Mimir observations, we were limited in our polarization accuracy by photon statistics, not systematic errors as was the case on the IRTF.

For NSFCAM2, position angle and efficiency calibration were measured using observations of S1 in $\rho$ Oph and comparing with our previous polarimetric results using the original NSFCAM. We used $P=3.90\%$ at $\theta = 28\degr$ for the intrinsic polarization of S1 in the $H$ band \citep{wil80}. We measured $P=3.78\%$ with NSFCAM2, which in principle would correspond to an efficiency of $97\%$. However, given our systematic error uncertainty of $\pm 0.3\%$ and the fact that all of the polarization optics are the same in NSFCAM2 and the original NSFCAM (which required no efficiency correction), we chose not to make any efficiency correction. Comparing the computed position angle of the comet (using S1 as the position angle calibrator) with the expected position angle derived from the Sun-Comet-Earth geometry at the epoch of our observations, produced agreement to within $3\degr$ or better, consistent with our systematic error.

For the Mimir observations we used the mean efficiency given in the instruments data reduction manual $(91\%)$ and made no position angle calibration. The instrumental position angle correction in Mimir is a function of location in the focal plane. We used S1 in $\rho$ Oph for as a check on the efficiency of Mimir in polarimetry mode, but did not make multiple observations at different locations in the focal plane. Aperture polarimetry of SW3 was performed by using a $3\arcsec$ synthetic aperture on our polarization images for both the optical and infrared observations. SW3 was relatively faint in 2006 April and we were not able to extract spatial information from our $H$ band observations using Mimir. Our optical polarimetry was limited by seeing and instrumental blurring at the $2\arcsec$ level, preventing the investigation of optical spatial variations. Only the $H$ band observations in 2006 May with NSFCAM2 had both the signal-to-noise and the spatial resolution (seeing was $0\farcs7$ FWHM) to allow investigation of spatial variations in the linear polarization. The results for the synthetic aperture polarimetry are listed in Table 2. 

$H$ band intensity images on May 12 and 13 are shown in Figures 1 and 2. The seeing was $0\farcs7$, which corresponds to $40$~km at Segment C on May 12 and $35$~km at Segment B on May 13. The inner comae of the comet segments are extended at our spatial resolution, although we can not resolve the myriad of small clumps that make up the southwest section of Segment B. The comae are extended most prominently in the direction away from the Sun (the tail), suggesting the release activity is distributed around the nucleus and not dominated by a single strong jet. For typical ejection velocities of $0.1-1.0$ km$~$s$^{-1}$, we are observing features formed on $1-10$ minute time scales.

\section{Results and Discussion} 

Our observations contain optical and near infrared polarimetric information for the coma as a whole at all epochs and more detailed polarimetric and photometric images at $H$ in 2006 May. In this section we first discuss the polarimetric observations, including the wavelength dependence and spatial variation of the fractional polarization. Changes in the fractional polarization with distance from the nucelus is an indicator of changes in the structure of the dust particles while in flight. Second, we discuss the observed surface brightness, in particular one dimensional cuts through the images for comparison with model calculations. We do find evidence that dust fragmentation is significant and this must be taken into account in any models of the polarization of scattered light in the comae of comets. Scattering models must reproduce a fractional polarization that remains nearly constant with wavelength from $\sim 0.4-2.2\micron$, but changes with location, even though the dust is undergoing significant fragmentation. In this paper no attempt is made to simultaneously model both the polarization properties of the comet dust and the evolution of the dust scattering properties with time. 

\subsection {Polarization}

$I~(0.87~\micron)$ and $H~(1.65~\micron)$ band synthetic aperture polarimetry values for Segment C and the NE component of Segment B are plotted versus phase in Figure 3. The magnitude of the polarization is typical for a high-polarization comet \citep[e.g.,][]{kel04}. The curved solid line in Figure 3 represents the mean R band polarization for a compilation of high polarization comets and the dashed line represents the trend for low polarization comets \citep{lev96}. The fact that our $I$ band data at smaller phase angles are $1-2\%$ higher than this trend most likely results from the longer wavelengths of our observations for the case of red polarimetric colors in optical red region of the spectrum. At larger phase angles, the average trend is not very well determined, but our observations are slightly below the $R$ band trend. Comets typically show a small increase in polarization with wavelength across the optical. \citet{lev96} argue that comets can be divided into two classes, ones with high polarization (the curved solid line in Figure 3) and ones with lower polarization at phase angles larger than $\sim40\degr$ (the dashed line in Figure 3). We would expect comet SW3 to be a low-polarization comet, as are the majority of Jupiter-family comets \citep{kol07}. The division into two optical polarization classes is probably due to dilution by molecular gas emission in the broadband filters often used at optical wavelengths \citep{kis01, kis04, jew04, joc05}. Polarimetry of 2P/Encke, a gas-rich, Jupiter-family comet, approaches high optical polarization values in small apertures \citep{jew04, joc05} and in narrower bandpasses chosen to avoid molecular emission bands. This is interpreted as being due to the dust and gas having different nucleocentric surface brightness profiles \citep{kol07}. If this interpretation is correct, all comets should show high polarization at near-infrared wavelengths at large phase angles, since molecular emission bands are much weaker at these wavelengths.

There is a correlation between comet polarization type and $10~\micron$ silicate emission. Dust-rich comets have a stronger $10~\micron$ silicate emission feature than gas-rich comets \citep{lev96, kol07}. \citet{kol07} present strong evidence that the two groups of polarimetrically different comets result from different evolution of the comets. New comets and periodic comets with large semi-major axes are characterized by high polarization at large phase angles since their dust is dominated by rather pristine comet material consisting of porous aggregates. Such porous particles can be easily accelerated by gas flow and reach large distances from the nucleus. Thus, the polarization of these comets does not depend on aperture size. Comets with smaller semimajor axes exhibit low polarization since their dust contains large compact particles formed by the highly processed surface material. Such compact particles have a tendency to concentrate close to the nucleus. Further out from the nucleus, the concentration of dust particles drops and the values of polarization becomes more strongly affected by gas contamination, decreasing the average value of polarization. Note that dust grain models by \citet{kim03} suggest that the polarization properties of these two types of grain aggregates should be similar in the absence of diluting gas emission. The high polarization of SW3 is indicative of porous aggregates, perhaps combined with the effects of fragmentation, producing grains easily accelerated to large distances from the nucleus.

SW3 has strong silicate emission \citep{har06} and high polarization both in the optical at $I$ and in the infrared at $H$ at high phase angles and at all distances from the nucleus we could measure (we define strong silicate emission as a $10~\micron$ spectral feature lying at least 20\% above the coninuum). One explanation for the unusually high polarization of SW3 well off the nucleus and into the tail is breakup of the nucleus and the subsequent release of unprocessed material that resembles the dust continuously being released in other highly polarized comets such as 1P/Halley and C/1995 01 (Hale-Bopp). This effect has been seen in other disintegrating comets \citep[e.g.,][]{kis02}. 

Another important characteristic of the polarization of SW3 is the change with wavelength (polarimetric color). Most previous infrared polarimetry, such as \citet{jg}, \citet{has97}, and \citet{kel04}, could not be easily compared to optical polarimetry since the optical and infrared polarimetry was not simultaneous and required extrapolation in phase angle. Our new results on SW3 are the among the few nearly simultaneous optical $(\lambda < 1\micron)$ and infrared $(\lambda > 1\micron)$ polarimetry observations of a comet \citep[see Table 3.5 in][]{kel06}. The polarimetric color for both segments is red (increases with the wavelength, $\frac{\delta P}{\delta \lambda}=+1.2\pm 0.4$) at $30-40\degr$ phase angles when comparing the $I$ and $H$ band observations. Red polarimetric color is typical for comets and most likely indicates the presence of porous aggregates \citep{kol04b}. However, there is some evidence for a more neutral, or slightly blue polarization color at $80-90\degr$ phase angles, at least for Segment C. To date, a blue polarimetric color has been observed only three times:
\begin{itemize} 
\item{} in comet 21P/Giacobinni-Zinner \citep{kis00}, a blue polarimetric color was explained by large abundance of organic materials or by large dust particles;
\item{} for comet Hale-Bopp \citep{kel04}, the polarimetric color was blue in the infrared although red in optical. However, this result is uncertain due to the need to extrapolate the observations in time and phase angle;
\item{} for comet 9P/Tempel-1, a blue polarimetric color was observed right after Deep Impact \citep{har07}. This blue polarimetric color was explained by presence of large amount of organics or ice in the Deep Impact ejecta \citep{har07}.
\end{itemize}
There is little comet polarimetry extending beyond $1~\micron$, but our data for SW3 and previous work on Hale-Bopp \citep{kel05} suggest that a blue polarimetric color could be common at larger phase angles. Possibly the blue polarimetric color is evidence of unprocessed, perhaps organic-rich comet material. In 2006 May we imaged areas close to the nucleus of SW3, where one would expect to find more pristine, recently released material, similar to that observed by Deep Impact. We note that blue polarimetric color was observed in Tempel 1 right after the impact and it turned red some days later \citep{har07}.

In Figure 4, we plot a cut through the intensity (Fig. 2) and polarization maps of Segment C in the Sun-Comet-Tail direction through the nucleus. The polarization shows a small decrease at the nucleus and rises outward from there. This is not unusual in comets \citep{kol01}, but in SW3 we are seeing changes on $50$~km scales, corresponding to $1-10$ minute time scales. In most imaging polarimetry of comets, the polarization rises with radial distance from the nucleus on much longer time and distance scales of order 1 hour and $1000-5000$~km. Thus, the reasons for the near-nucleus change in polarization for SW3 and other observed comets may not be the same. The cause of the increase in polarization in SW3 with distance from the nucleus might be fragmentation of particles (since polarization usually gets higher as the particle size decreases) or a change in refractive index due to evaporation of volatiles. Our $I$ band polarization maps are not of sufficient quality to combine with the $H$ band observations, and without a map of polarimetric color, it will be difficult to distinguish between these two \citep{kol01}.

Segment B (Figure 5) displays a similar overall level of polarization, but is broken into two sections aligned with the Sun-Comet-Tail direction and only $3\arcsec$ apart. This makes it impossible to distinguish changes in polarization into the tail since the tail from the NE fragment overlaps the SW fragment. In the direction of the Sun from the nucleus of the NE component of Segment B, the polarization shows a distinct drop in strength from 25\% to 22\%. The entire tail complex is consistent with a constant polarization of $\sim 25$\%. The drop in polarization in sunward direction may indicate the presence of particles too large to be quickly moved into the tail direction by radiation pressure, although why this would be the case for Sement B and not Segment C is not clear.

\subsection{Photometry}

Steady, isotropic dust outflow in the absence of a strong external force, produces a dust grain coma density that varies with $r^{-2}$, where $r$ is the distance from the comet nucleus.  Integrating a $r^{-2}$ coma density profile along an observer's line-of-sight yields a $\rho^{-1}$ surface brightness profile, where $\rho$ indicates radial offset distance projected on the sky.  Deviations from this profile indicates deviation from steady, isotropic outflow or some change in grain scattering characteristics with time since release.

A plot of $H$ band intensity along the Sun-Comet-Tail direction through Segment C is shown in Figure 6. The profile of a star and a $\rho^{-1}$ power law are shown for comparison. The surface brightness of the coma for Segment C decreases with radius more slowly than $\rho^{-1}$ out to about $1\arcsec$ for the Sunward side and out to about $3\arcsec$ into the tail. In the following sections, for both Segments B and C, we consider three possibilities to explain the surface brightness profiles: 1) gas-dust outflow coupling, 2) solar radiation pressure, and 3) grain fragmentation. 

\subsubsection{Gas-dust coupling}
Gas and dust outflows decouple within several radii from the
surface of comet nuclei \citep{com97}.  Before fragmentation in 1995,
visual observations of comet 73P by \citet{boe99} constrained
the radius to be $\lesssim1.1$~km, and the 2006 fragments are certainly
smaller. \citet{tot06}  and \citet{wea06} present  HST observations of the nucleus of fragment C during its close approach to Earth in 2006 and estimate a radius of $0.41\pm0.02$~km. Since the scale for gas and dust decoupling is only a few times greater than the radius of the nucleus, the decoupling will take place within a few km from the nucleus. Our NSFCAM2 polarimetry probes the coma on $40$~km scale lengths, well beyond the region of gas and dust decoupling.  Moreover, dust grains accelerated by the gas outflow will cause the $\rho^{-1}$ coma density profile
to steepen, not flatten as seen in the inner portion of Fig.~6.  Once decoupled from the
gas, the dust grains will resume a steady, isotropic outflow.  The
effects of gas-dust outflow coupling near the nucleus surface are
wholly unresolved in our observations.

\subsubsection{Radiation pressure}
Solar radiation pressure will also modify observed coma profiles.
To investigate this effect, we simulated the coma of fragment C using a
dynamical model for comet dust \citep{kel06}.  The model accounts
for both solar radiation pressure and the force of gravity from the
sun and planets acting on the dust grains.  The model's synthetic
imager has been upgraded to include a simple description of scattering by spherical grains.  In
this description, the scattering efficiency varies as $(2\pi a /
\lambda^4)$ for $a < \lambda / 2\pi$ and $2\pi a / \lambda$ for $a >
\lambda / 2\pi$, where $a$ is the grain radius, and $\lambda$ is the
wavelength of scattered light.  The model parameterizes grains with the ratio $\beta =
F_{rad} / F_{g}$, where $F_{rad}$ is the force of solar radiation and
$F_{g}$ is the force of gravity.  Since both forces vary with solar distance $r_h$ as
$r_{h}^{-2}$, $\beta$ is constant for a given grain size and mass.  The
ratio reduces to $\beta = Q_{pr} 0.57 / a \rho_d$, where $Q_{pr}$ is the
efficiency of radiation pressure (assumed to be $\approx 1$) and
$\rho_d$ is the grain density.  We treat grains as low density spheres
with $\rho_d = 1$ gm cm$^{-3}$.  We use grains with $0.001 \leq \beta \leq 1$,
approximately corresponding to $600~\micron{} \geq a \geq
0.6~\micron{}$, a dust production rate proportional to the visual coma
light curve near perihelion, $Q_d \propto r_{h}^{-2.6}$
\citep{yoshida07}, and a particle size distribution similar to that
measured by the \textit{Stardust} spacecraft in the coma of 81P/Wild, with
$dn/da \propto a^{-3.25}$ \citep{gre04}.  The oldest grains tracked
were 20--30 days old. The synthetic images are computed corresponding to the geometry of the comet ejecta as viewed from the Earth on the days of our IRTF observations (May 11-13 2006).

We simulated the coma (for comparison with the $H$ band image) of fragment C with two sets of ejection velocity
parameters.  The first set (Method 1) ejected $10^7$ grains with
$v_{ej} = v_0 \sqrt{\beta / r_h}$, where $v_0$ is the ejection
velocity of $\beta = 1$ grains at $r_h = 1$~AU.  We executed the
simulation with different values for $v_0$ of 0.02, 0.05, 0.1, 0.2, and $0.5$ km$~$s$^{-1}$. In this scenario, small grains with large values of $\beta$ have faster ejection velocities than the larger grains. The subsequent interaction of all the ejected grains with solar radiation pressure is computed in small time steps and the trajectory of each grains is tracked. At any moment in time, the position of all the model grains can be used to compute a synthetic surface brightness map by projecting the grain positions onto the plane of the sky and counting the grains (weighted by their individual scattering cross section) in each synthetic pixel.

The second set of ejection parameters (Method 2) picked grains with ejection velocities
independent of $\beta$, where $v_{ej} \leq v_0 \sqrt{1 / r_h}$.  We
chose the same set of $v_0$ values as in Method 1.  Due to the large
range in ejection velocities, many more test grains were required in
the second simulation.  We chose a strategy that would produce $\approx 10^6$
grains per $\beta$ decade in the IRTF field-of-view (many of the
$10^7$ grains in Method 1 are outside of the IRTF field-of-view). In this scenario small grains with large values of $\beta$ can have a wide range of velocities, in particular, they can be ejected with lower velocities than in Method 1. 

Figure 7 presents the simulated images of fragment C
from Method 1, smoothed with a Gaussian kernel (FWHM = $0\farcs7$).
The $v_0 = 0.02$ km$~$s$^{-1}$ simulation produces a projected distribution shaped like a tail, but as
the velocity is increased to $v_0 = 0.1$ km$~$s$^{-1}$, the tail evolves
into an azimuthally symmetric coma.  Circular isophotes terminated
at a paraboloid of revolution, with the focus in the Sun direction, is typical of dust emission from an
isotropic point source when a 1-to-1-to-1 mapping of $\beta$, size,
and $v_{ej}$ is employed \citep{combi94}. Grains with high $\beta$ (small size) are ejected at higher velocities, but also are also more strongly accelerated by solar radiation pressure. Comparing the images to
Figure 2, we find that no model image reproduces the two key
features of the observed coma: 1) a dust coma with the observed large angular extent, and 2) ellipsoidal isophotes that also extend in the anti-solar direction.  We deduce that either there are strong
asymmetries in the dust ejection and production that an isotropic
model (by definition) does not reproduce, or there is a distribution
of grain velocities independent of $\beta$ in the coma.  Simulating strong ejection
asymmetries is beyond the scope of our dynamical model.  Distributions
of velocities for a given $\beta$-value are treated with ejection
Method 2.

Figure 8 presents the simulated images of fragment C
from Method 2 smoothed with a gaussian kernel (FWHM = $0\farcs7$).  In
contrast to Method 1, a sharp tail exits in all simulations.  This
sharp tail is comprised of grains ejected at very low velocities from
the nucleus.  Such a feature does not appear in our image of fragment
C; therefore, we modified Method 2 by removing low velocity grains by subtracting an image composed of all grains with a velocity less than some minimum $v \leq v_{min}$. Effectively, we are ejecting grains with a wide distribution of velocities but imposing a non-zero minimum grain velocity.  The isophotes
of the resultant simulated images are shown in Figure 9. They are more elliptical and better match
the observed isophotes of fragment C than Methods 1 or 2 when the minimum velocity is no less than $v_{min} = 0.005$ km$~$s$^{-1}$.

Dust acceleration by gas expansion is very efficient \citep{com97}
and a wide distribution of grain velocities for a given $\beta$-value
is not expected.  Alternatively, large grains may be ejected into the
coma with low velocities, subsequently fragmenting into smaller grains,
thus producing a population of small (high $\beta$) grains with low velocities.
\citet{combi94} studied grain fragmentation in a Monte Carlo model of
comet 1P/Halley's dust coma.  Elongated isophotes in images of comet
Halley, similar to those observed in our images of fragment C,
were well described by grain fragmentation.  Below, we treat grain
fragmentation with a simple model.

\subsubsection{Grain fragmentation}

Break up or fragmentation of comet dust particles has been discussed in the context of explaining an extended source of gas in some comets \citep[e.g.,][]{gre98}, and in the context of extended CO in comet Halley, although sublimation is considered as the main the reason for the extended source. Sublimation was considered as the explanation for the deviation of the Halley brightness profile from a simple 1/$\rho$ law by \citet{toz04}. Organics may hold grain aggregates together until the organics vaporize \citep{obe04, obe07}. In this section, we consider the effects of a simple breakup model for our images at in 2006 May, without reference to the actual mechanism of fragmentation. 

The model takes two input parameters: $\epsilon$, the probability of a breakup in a $1$~km distance, and $n$, the maximum number of breakups allowed. Each breakup results in two grain aggregates, each with half the volume (mass) of the parent and a combined scattering cross section $2^{\frac{1}{3}}$ larger. The total increase in scattering cross section would be $2^{\frac{n}{3}}$ at large enough radii for all $n$ breakups to have taken place. The total number of new grain aggregates will be $2^{n}$, each with radius $a=a_{0}$/$2^{\frac{n}{3}}$, where $a_0$ is the radius of the original parent. The parameter $\epsilon$ controls the physical distance scale over which breakup typically takes place.

Given the input parameters $\epsilon$ and $n$, the radial distance from the `nucleus' is stepped in very small increments and the probability of a breakup is computed at each step using a random number generator and the value for $\epsilon$. The model continues to step out in radius (stopping if $n$ has been reached) until a distance of $10^{4}$~km ($\sim200\arcsec$ in projection). At each step, the mass density of scatterers is decreased by $r^{-2}$, consistent with the assumption of constant velocity outflow. The calculation is repeated 1000 times to generate an average radial dependence of the scattering cross section. This radial profile is then integrated along lines of sight with a range of impact parameters to produce a model surface brightness profile. 

We assume the constant velocity region is well established within our seeing disk ($0\farcs7$ FWHM, $35-40$~km). We find that a typical value for $\epsilon$ that fits the observations is $0.03$~km, so the first breakup commonly takes place well beyond 1~km distance. At projected radii less than $0\farcs02$ ($\sim 1$~km), the model surface brightness is set at the value for $0\farcs02$, well within our seeing disk. The model profile is then convolved with a Gaussian with a FWHM of $0\farcs7$ and normalized to 1 at $r=\rho = 0$. We first compare our model results with the radial profiles of Segments C and B in the direction perpendicular to the Sun-Comet-Tail direction. We refer to these profiles as `crosscuts.'

Our goal with this model is to determine if a simple breakup scheme can explain the observed brightness profiles of the B and C segments. In particular, we are interested in determining the number of breakups necessary to create the observed departure from a $\rho^{-1}$ surface brightness. The observed polarization in the near-infrared is no higher than in the visual. Thus the grain aggregates can not fragment to the point where small, nearly Rayleigh particles (compared to a wavelength of $1.65~\micron$) are all that is left, or the polarization at $H$ band would be much higher than observed.

The general behavior of the model is illustrated in the two panels of Figure 10. In Figure 10a the value for $n$ is kept at 10 and the model results for three different values of $\epsilon$ are plotted. These profiles are compared with the profile of a star and the model results for the case with $n = 0$, labeled {\it No Breakup}. The model results for {\it No Breakup} show a seeing broadened central peak wider than the stellar profile due to the fact that the fragment has a $\rho^{-1}$ profile and is not a point source like a star. It quickly establishes a $\rho^{-1}$ profile beyond $\sim 1\arcsec$. Small values of $\epsilon$, corresponding to a longer average distance between breakups, produce a bump in the profile that is not seen in the data. Very small values of $\epsilon$ will produce a model profile similar to a limb brightened spherical shell sitting on top of the {\it No Breakup} profile at a correspondingly large angular distance from the nucleus. If $\epsilon = 0$, the model is equivalent to the {\it No Breakup} case, since no spherical shell is produced at any radius. Values of $\epsilon$ significantly larger than 0.03 cause the model profile to narrow and approach the case with {\it No Breakup}. In Figure 10b, the value for $\epsilon$ is kept constant at $\epsilon = 0.03$ and the model results for different values of $n$ are plotted. A number of breakups significantly exceeding 10 produces a radial profile much more extended than seen in the data. If the number of breakups is only 5, a narrower profile similar to the case for $\epsilon = 0.09,~~n=10$ results. 

Model fits to the observed profiles are shown in the three panels of Figure 11. In panel (a) we plot the observed surface brightness profile of a cut through the nucleus of the NE component of Segment B perpendicular to the Sun-Comet direction (crosscut). A model with $\epsilon = 0.03$ and $n=10$ fits the data within $\pm 10\%$ at all radii. Models with $n > 14$ and $n < 8$ depart from the data by more than 15\% for all values of $\epsilon$. At large radii the data are a factor of 4 above the model profile for {\it No Breakup}. Naively this would correspond to only 6 breakups ($2^{\frac{n}{3}} = 4$ when $n = 6$), not the 10 breakups found in the model fit with $\epsilon = 0.03$. However, there is significant breakup at scales comparable to our seeing limit in the model when $\epsilon \ge 0.025$. Because the smoothed model profile is normalized to 1 at the smallest radius, these early breakups are not resolved. This causes the final, normalized profile to not be as far below the data as expected, based only on the value of $n$. In Figure 11b we make the same plot for a crosscut through Segment C. The model plotted corresponds to $\epsilon = 0.03$ and $n = 12$, very similar results to our fit to Segment B. This would be expected as both segments probably have similar dust composition and structure.

Our model illustrates how significant the increase in surface brightness is over the case where there is no change in the grain scattering characteristics with distance from the nucleus. If fragmentation explains the observed surface brightness profiles, then a significant change in the mean grain aggregate size must take place. Our model results for the crosscut radial profiles suggest a reduction in mean grain aggregate radius of a factor of $\sim 10$ (i.e., $\frac{a_0}{a}=2^\frac{n}{3}\sim 10$ for $n=10$). 

For a cut along the Sun-Comet-Tail direction, our breakup model is less successful. Panel c in Figure 11 shows the radial surface brightness of the tail for Segment C (see Figure 6). The model plotted corresponds to $\epsilon = 0.02$ and $n = 15$, significantly different than the case for the crosscut (Figure 10b). The lower value for $\epsilon$ delays breakup enough to extend the profile out from the nucleus, but results in a bump in the model profile at $\sim 1.5\arcsec$. The large $n$ raises the overall level of the $\rho^{-1}$ portion. The fit is not very good, and we found that no combination of input parameters could produce the long, slow, smooth drop in surface brightness seen in the tail of Segment C. 

There is no obvious reason why dust released into the tail should have significantly different breakup characteristics than dust released perpendicular to the Tail. Likely the difference between the tail and crosscut profiles are due to dynamical effects, for example, slow moving, small grains (the daughters of fragmentation) are more likely to be pushed by the sun into the anti-sun direction. Indeed, the brightness profile of the Sunward side (Figure 6) shows a steep drop in intensity between $1-2\arcsec$, faster than $\rho^{-1}$, suggesting radiation pressure is effective in removing dust on $50-100$~km scales and sweeping it into the tail.  For example, the sunward surface brightness profiles for the images from our dynamical model method 2 (Figure 8) are all steeper than a $\rho^{-1}$ profile. It is beyond the scope of this paper to combine our dynamical and breakup models, and this will be left for future work.

\section{Summary}

We have presented nearly simultaneous optical-near infrared imaging polarimetry observations of Comet 73P/Schwassmann-Wachmann 3 fragments B and C in the $I$ and $H$ bandpasses at solar phase angles of approximately 35 and $85\degr$. Observations during the closest approach from 2006 May 11-13 achieved a spatial resolution of $35-40$~km in the coma. The level of polarization was typical for active comets, but higher than expected for a Jupiter family comet. The polarimetric color was slightly red at a phase angle of $\sim 35 \degr$ and either neutral or slightly blue at a phase angle of $\sim 85 \degr$. 

High quality images of SW3 in the $H$ band from the 2006 May close approach show segments clearly depart from a simple $1/\rho$ surface brightness profile for the first $50-200$~km from the nucleus. We built a dynamical model that requires a wide distribution of velocities (at least a factor of 10) for a given grain size to be present in the coma in order to approximate the ellipsoidal shape of the observed fragment C isophotes. Our simulations of grain aggregate breakup and fragmentation are able to reproduce the observed profile perpendicular to the Sun-Comet axis, but produced poorer fits to the observations along this axis (into the tail). According to our breakup model, the amount of required fragmentation is significant, with a reduction in the mean grain aggregate size by about a factor of 10 taking place between 2 and $200$~km from the nucleus. Most likely the morphology of SW3 is due to a combination of the effects of radiation pressure and dust aggregate breakup, but more detailed modeling will be necessary.

\section{Acknowledgments}

We would like to thank Eric Tollstrup and the support staff for the IRTF for helping with NSFCAM2 in its first use as a polarimeter. We also wish to thank Gerry Ruch, Andrew Helton, and Martha Boyer for their assistance with the optical polarimetry. This research was conducted in part using the Mimir instrument, jointly developed at Boston University and Lowell Observatory and supported by NASA, NSF, and the W.M. Keck Foundation. This research has been supported in part by National Science Foundation grants AST-037446, AST-0705030, AST-0440936, AST-0706980, and AST-0607500.

\begin{deluxetable}{ccccc}
\tablenum{1}
\tablewidth{0pt}
\tablecaption{Observing Log}
\tablehead{
\colhead{Date (2006)} & \colhead{UT (Hr)} & \colhead{Telescope} & \colhead{Instrument} & \colhead{Filter}
}
\startdata

April 18 & 3:35 & MLOF & OPTIPOL & $I$ \\
April 18 & 9:30 & Perkins & Mimir & $H$ \\
April 19 & 5:41 & MLOF & OPTIPOL & $I$ \\
April 19 & 8:15 & Perkins & Mimir & $H$ \\
May 11 & 9:30 & MLOF & OPTIPOL & $I$ \\
May 12 & 10:08 & MLOF & OPTIPOL & $I$ \\
May 12 & 13:32 & IRTF & NSFCAM2 & $H$ \\
May 13 & 8:26 & MLOF & OPTIPOL & $I$ \\
May 13 & 13:26 & IRTF & NSFCAM2 & $H$ \\

\enddata
\end{deluxetable}

\begin{deluxetable}{cccccccc}
\tablenum{2}
\tablewidth{0pt}
\tablecaption{Aperture Polarimetry $(3\arcsec)$}
\tablehead{
\colhead{Date (2006)} & \colhead{UT (Hr)} & \colhead{Segment} & Phase Angle & \colhead{Filter} & \colhead{P(\%)} &\colhead{$\epsilon P(\%)$} & \colhead{$\Delta$ (AU)} }
\startdata

April 18 & 3:35 & C & 35.8 & $I$ & 6.0 & 0.2 & 0.227 \\
April 18 & 9:03 & C & 36.0 & $H$ & 7.0 & 0.3 & 0.225 \\
April 19 & 5:41 & B & 35.1 & $I$ & 5.5 & 0.2 & 0.233 \\
April 19 & 8:15 & B & 35.6 & $H$ & 6.5 & 0.3 & 0.232 \\
May 11 & 9:30 & B & 78.2 & $I$ & 22.7 & 0.2 & 0.071 \\
May 12 & 10:08 & C & 87.3 & $I$ & 24.4 & 0.2 & 0.079 \\
May 12 & 13:32 & C & 87.9 & $H$ & 23.5 & 0.3 & 0.079 \\
May 12 & 8:26 & B & 82.9 & $I$ & 23.8 & 0.2 & 0.069\\
May 13 & 13:26 & B & 88.6 & $H$ & 24.5 & 0.3 & 0.067 \\

\enddata
\end{deluxetable}

\setcounter{figure}{0}
\begin{figure}
\plotone{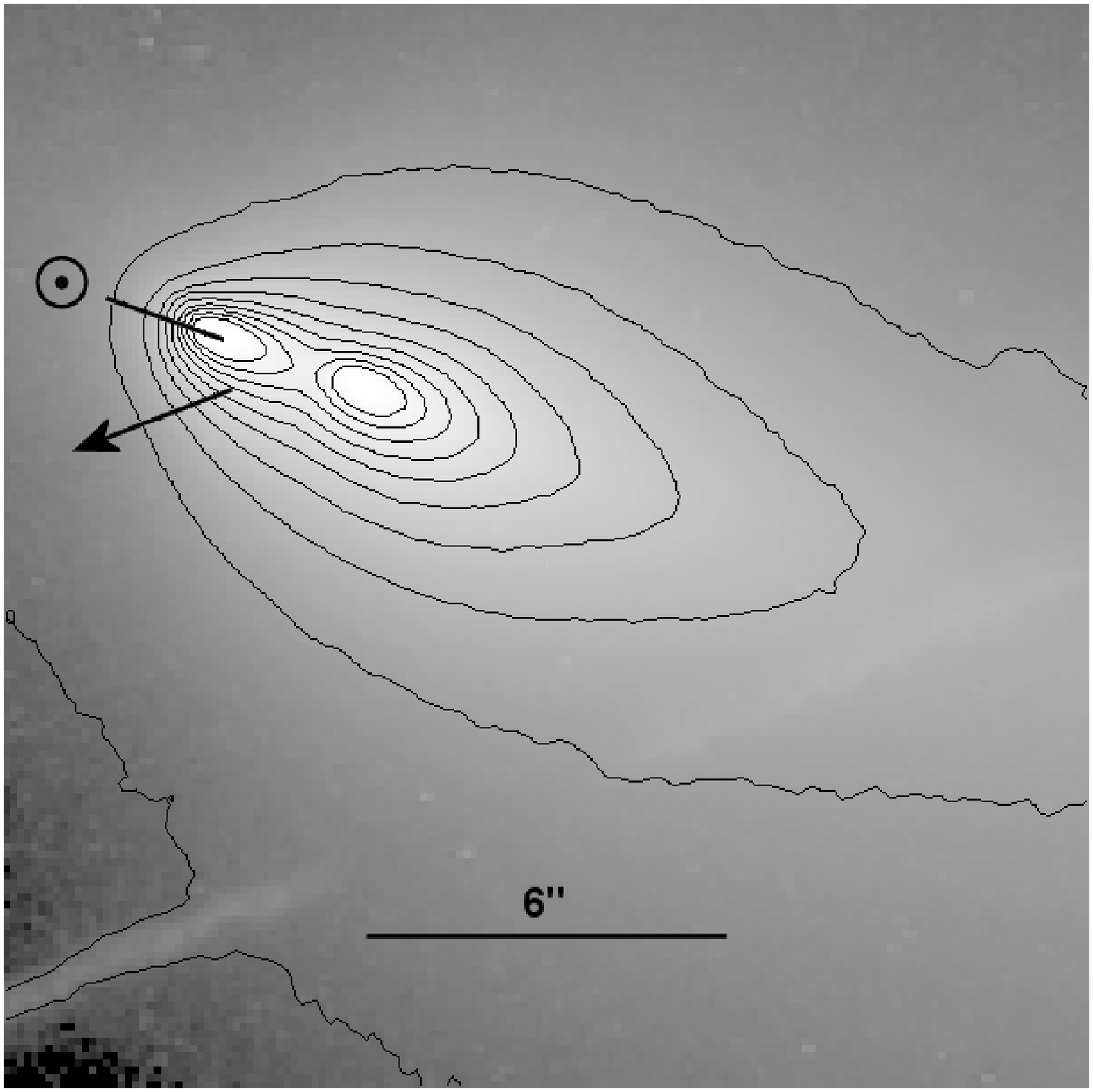}
\caption{$H$ band image of segment B of Comet 73P/Schwassmann-Wachmann 3 on 2006 May 13. The contours are linear, spaced $8.75\%$ of the peak intensity apart, and the lowest contour is $4\%$ of the peak. The arrow indicates the direction of motion of the comet on the sky. The direction of the Sun is indicated by the solid line and this is the direction of the Sun-Comet-Tail cuts discussed in the text. North is up and East is on the left.}
\end{figure}

\begin{figure}
\plotone{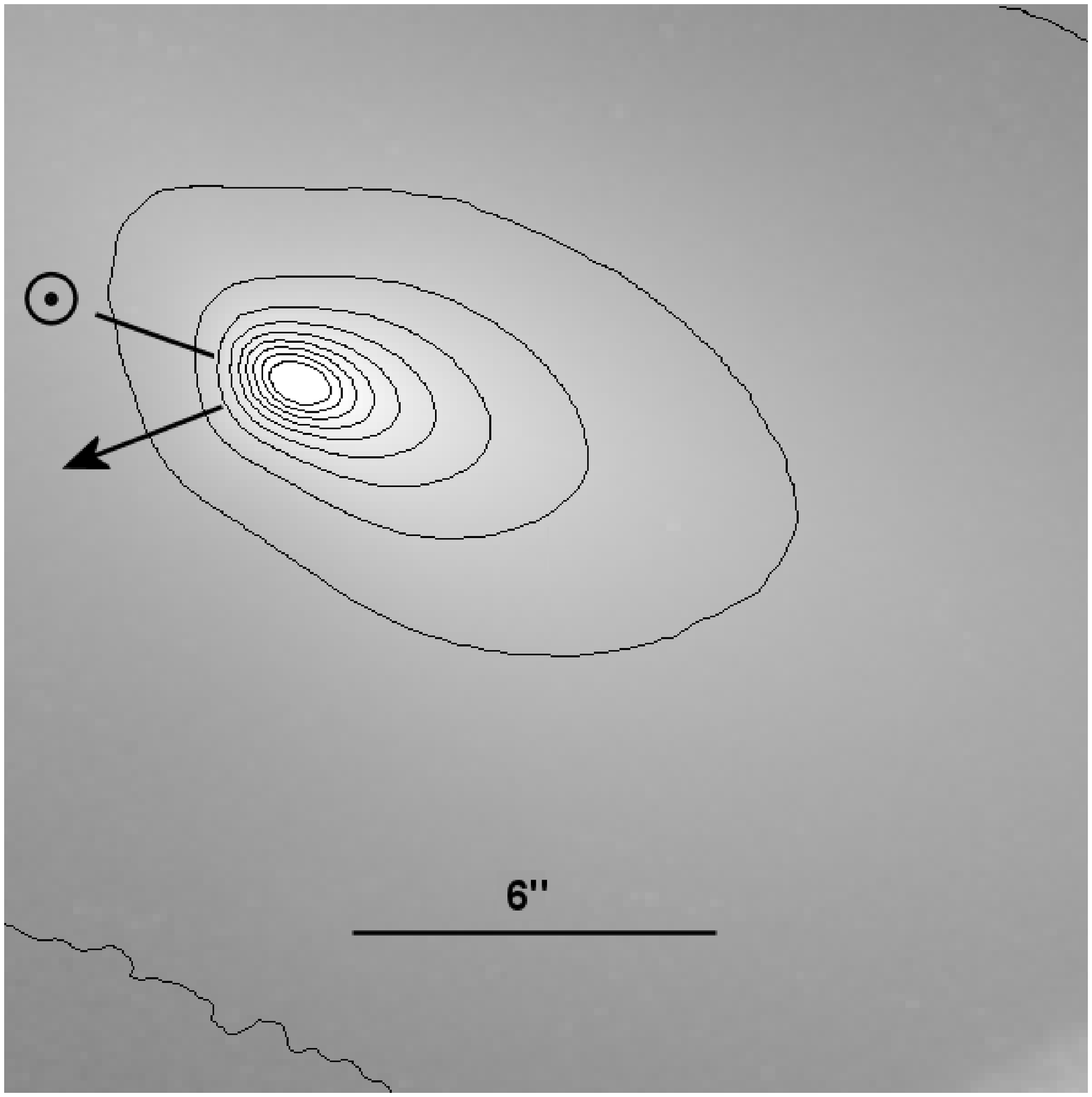}
\caption{$H$ band image of segment C of Comet 73P/Schwassmann-Wachmann 3 on 2006 May 12. The contours are linear, spaced $8.5\%$ of the peak intensity apart, and the lowest contour is $3.4\%$ of the peak. The arrow indicates the direction of motion of the comet on the sky. The direction of the Sun is indicated by the solid line and this is the direction of the Sun-Comet-Tail cuts discussed in the text. North is up and East is on the left.}
\end{figure}

\begin{figure}
\plotone{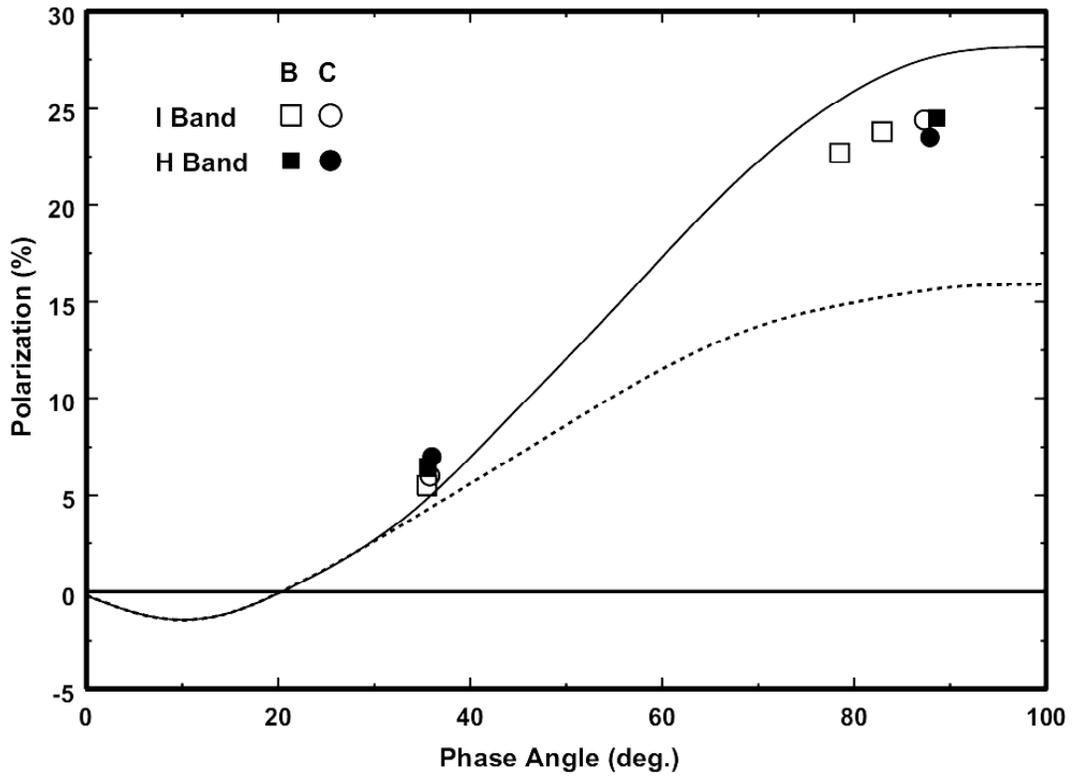}
\caption{Plot of linear polarization against phase angle for the aperture polarimetry listed in Table 2. The curved solid line is the general trend for high polarization comets in the optical at R band and the dashed line is for low polarization comets \citep{lev96}. The horizontal line delineates zero fractional polarization.}
\end{figure}

\begin{figure}
\plotone{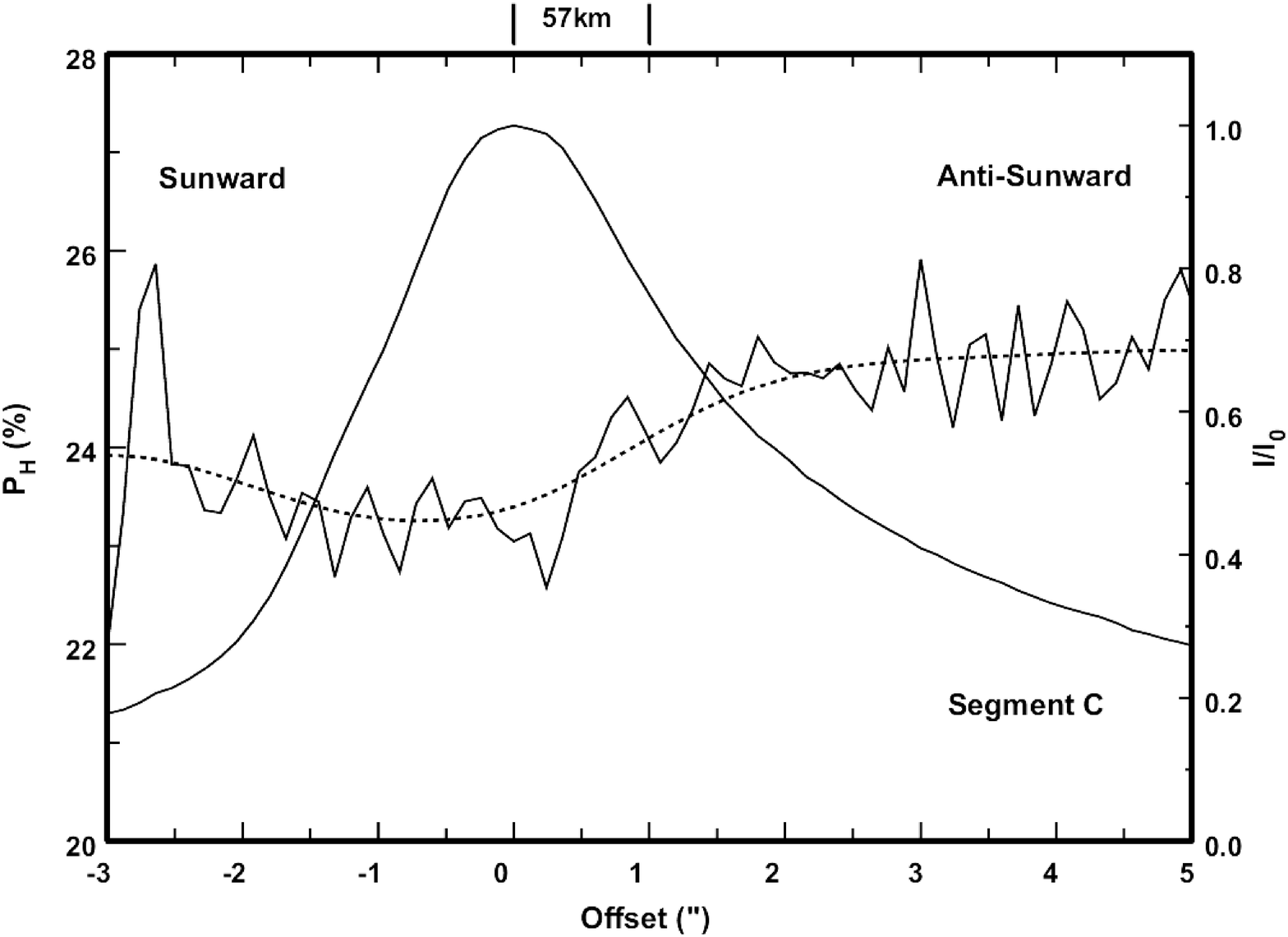}
\caption{Plot of normalized intensity (smooth solid line) and linear polarization (jagged solid line) in the $H$ band against offset angle along a cut aligned with the Sun-Comet-Tail direction on the sky for Segment C at 13:32 UT, 2006 May 12 (see Fig. 2). The dashed line is a smoothing spline fit to the linear polarization. Note that the Y axis starts at 20\% polarization.}
\end{figure}

\begin{figure}
\plotone{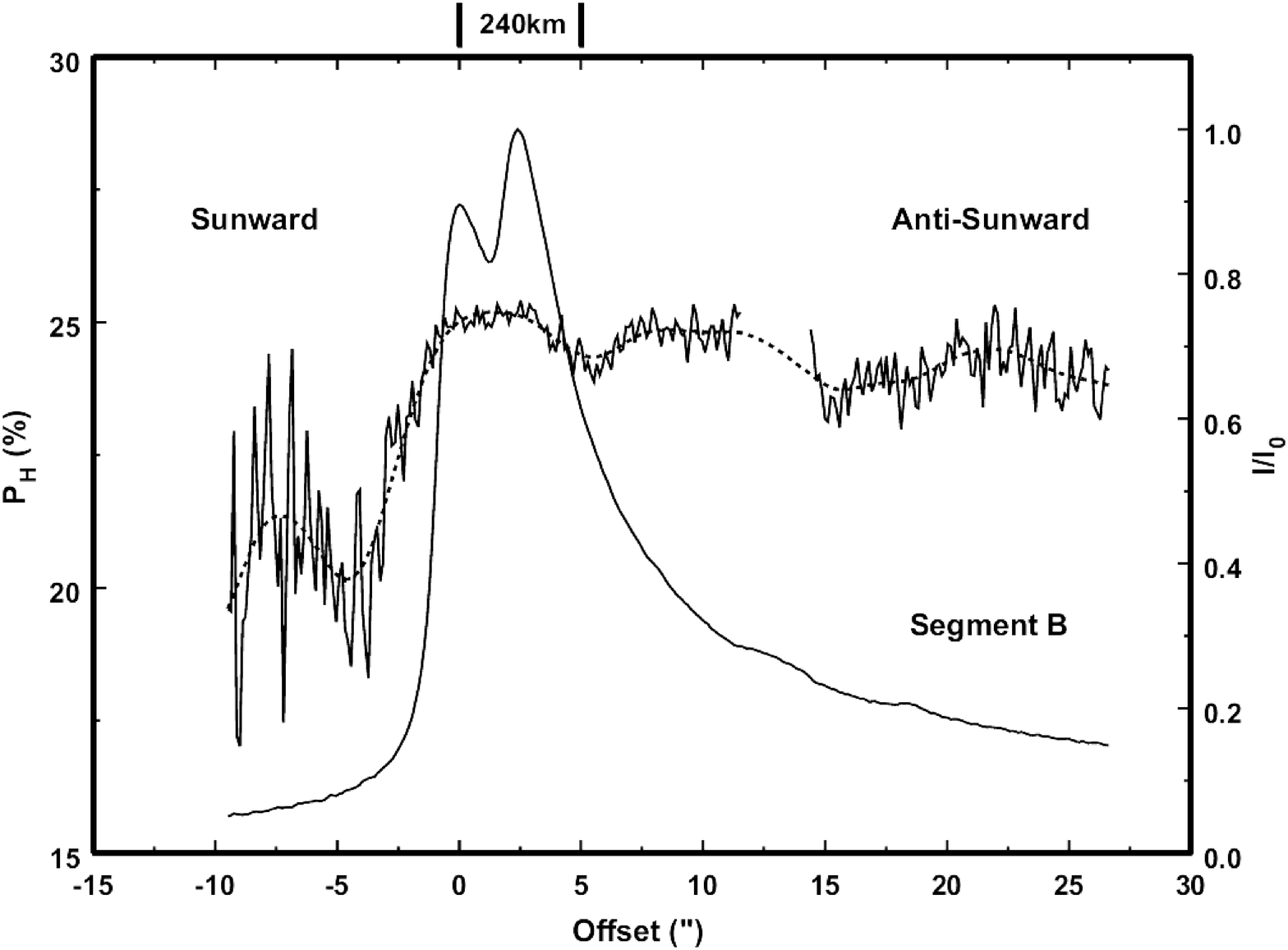}
\caption{Plot of normalized intensity (smooth solid line) and linear polarization (jagged solid line) in the $H$ band against offset angle along a cut aligned with the Sun-Comet-Tail direction on the sky for Segment B at 13:26 UT, 2006 May 13 (see Fig. 1). The dashed line is a smoothing spline fit to the linear polarization. Note that the Y axis starts at 15\% polarization.}
\end{figure}

\begin{figure}
\plotone{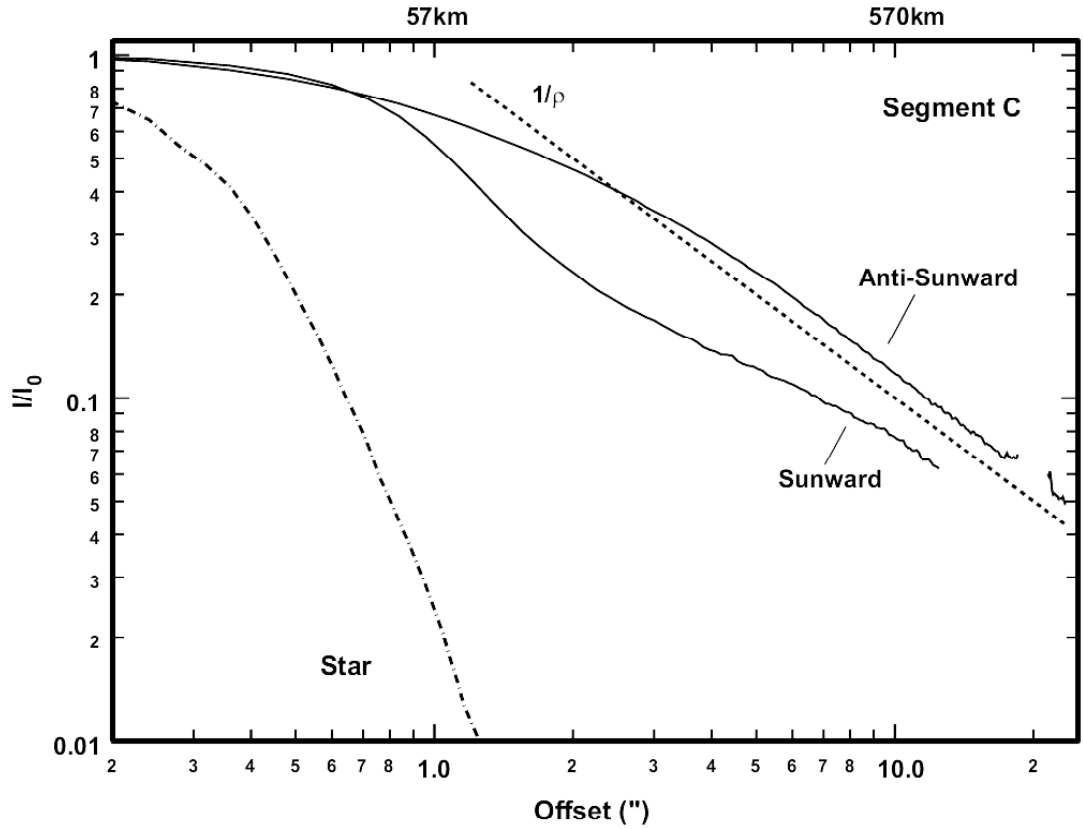}
\caption{Log-log plot of normalized intensity against offset angle from the nucleus for Segment C in the $H$ band along the Sun-Comet-Tail direction on the sky (see Fig. 2). The dashed line represents a $1/\rho$ dependence. The seeing was $0\farcs7$ FWHM, as indicated by the profile of a star (dot-dashed line).}
\end{figure}

\begin{figure}
  \plotone{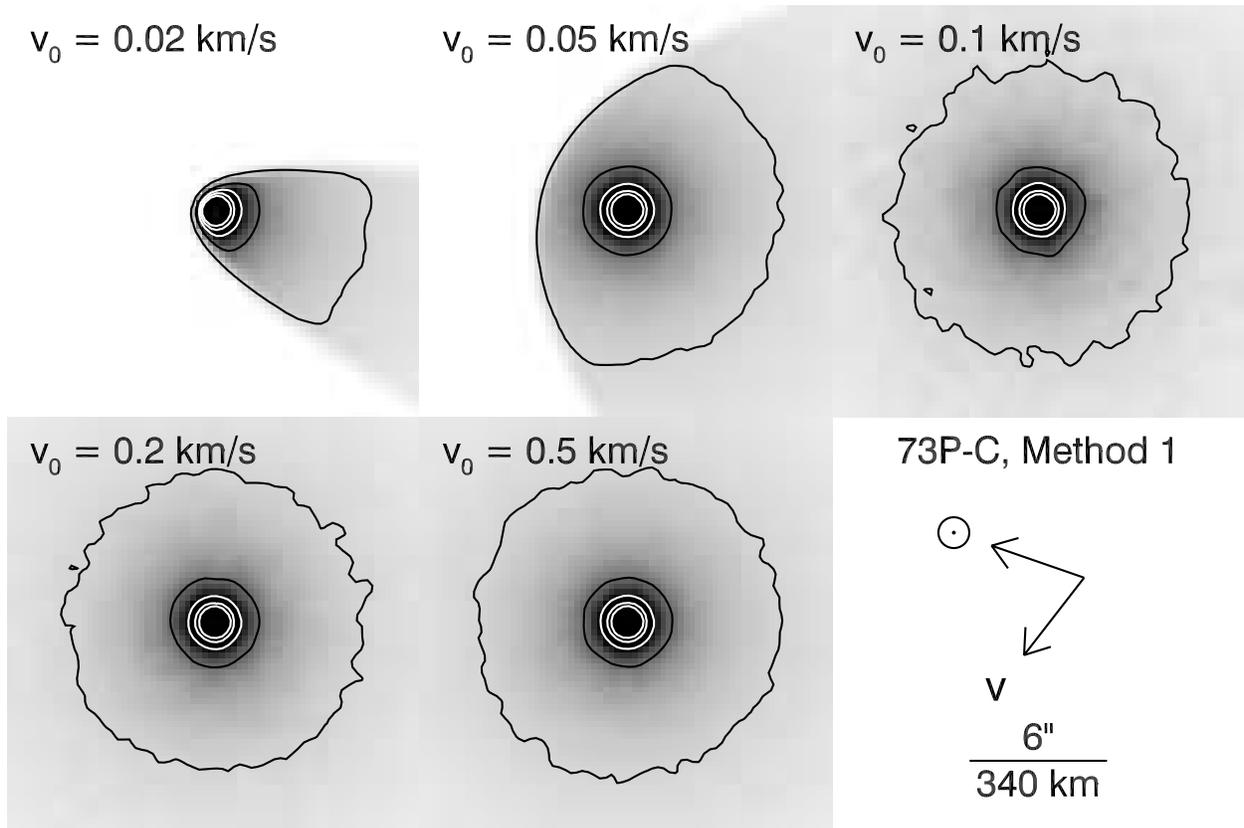}
  \caption{Simulated images of fragment C, where $v_{ej} = v_0
    \sqrt{\beta/R}$ km$~$s$^{-1}$ (Method 1; see text).  Contours
    are linear, spaced 8.5\% of the peak intensity apart, and the
    lowest contour is 3.4\% of the peak (same as Figure 2). These
    images have circular isophotes truncated at a paraboloid of
    revolution, a result of the 1-to-1-to-1 mapping of grain $\beta$,
    size, and $v_{ej}$ in the model \citep{combi94}, and do not
    resemble the observed isophotes in Figure 2.}
\end{figure}

\begin{figure}
  \plotone{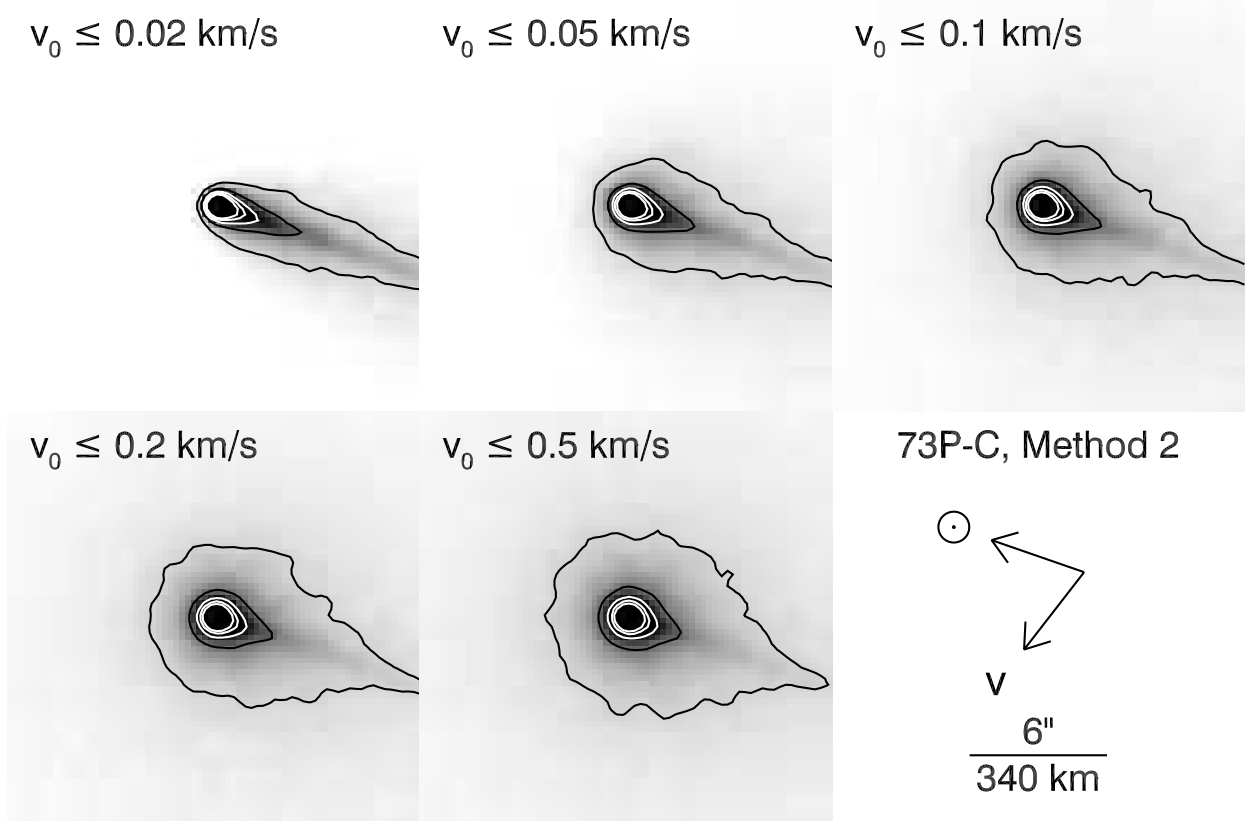}
  \caption{Simulated images of fragment C where $v_{ej} \leq v_0
    \sqrt{1/R}$ km$~$s$^{-1}$ (Method 2; see text).  Contours
    are linear, spaced 8.5\% of the peak intensity apart, and the
    lowest contour is 3.4\% of the peak (same as Figure 2).  These
    images have stronger isophote warping in the anti-sun direction, a
    better match to the ellipsoidal isophotes of our observation of
    fragment C (Figure 2).  The narrow tail is comprised of very
    low ejection velocity grains ($v_{ej} \approx 0.0$ km$~$s$^{-1}$)
    and is not present in our observations.}
\end{figure}

\begin{figure}
  \plotone{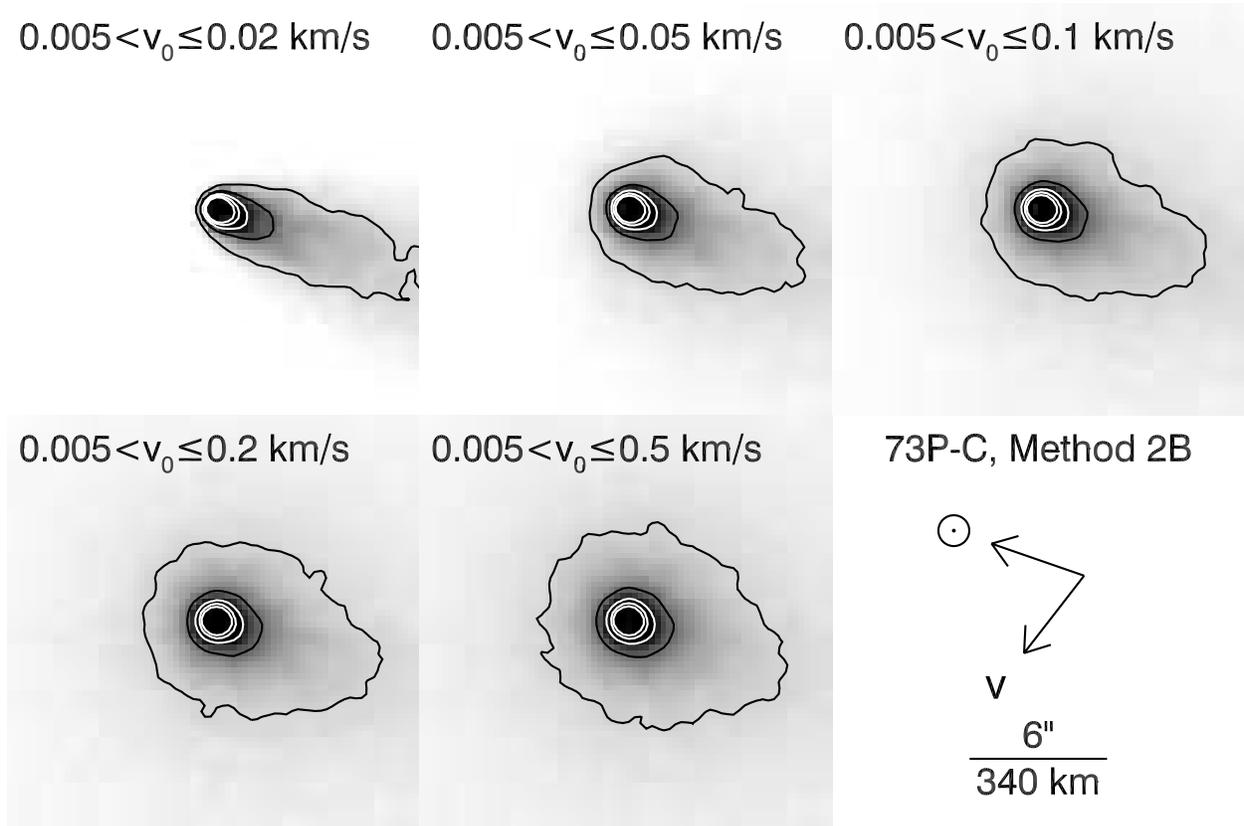}
  \caption{Simulated images of fragment C where $v_{ej} = v_0 \sqrt{1.0/r_h}$ km$~$s$^{-1}$ (modified Method 2, see text).
    Contours are linear, spaced 8.5\% of the peak intensity apart, and
    the lowest contour is 3.4\% of the peak (same as Figure 2).
    The chosen velocity distributions remove the narrow tail in
    Figure 8 and show better agreement with our observation of
    fragment C (see Fig. 2)}
\end{figure}

\begin{figure}
\epsscale{0.5}
\plotone{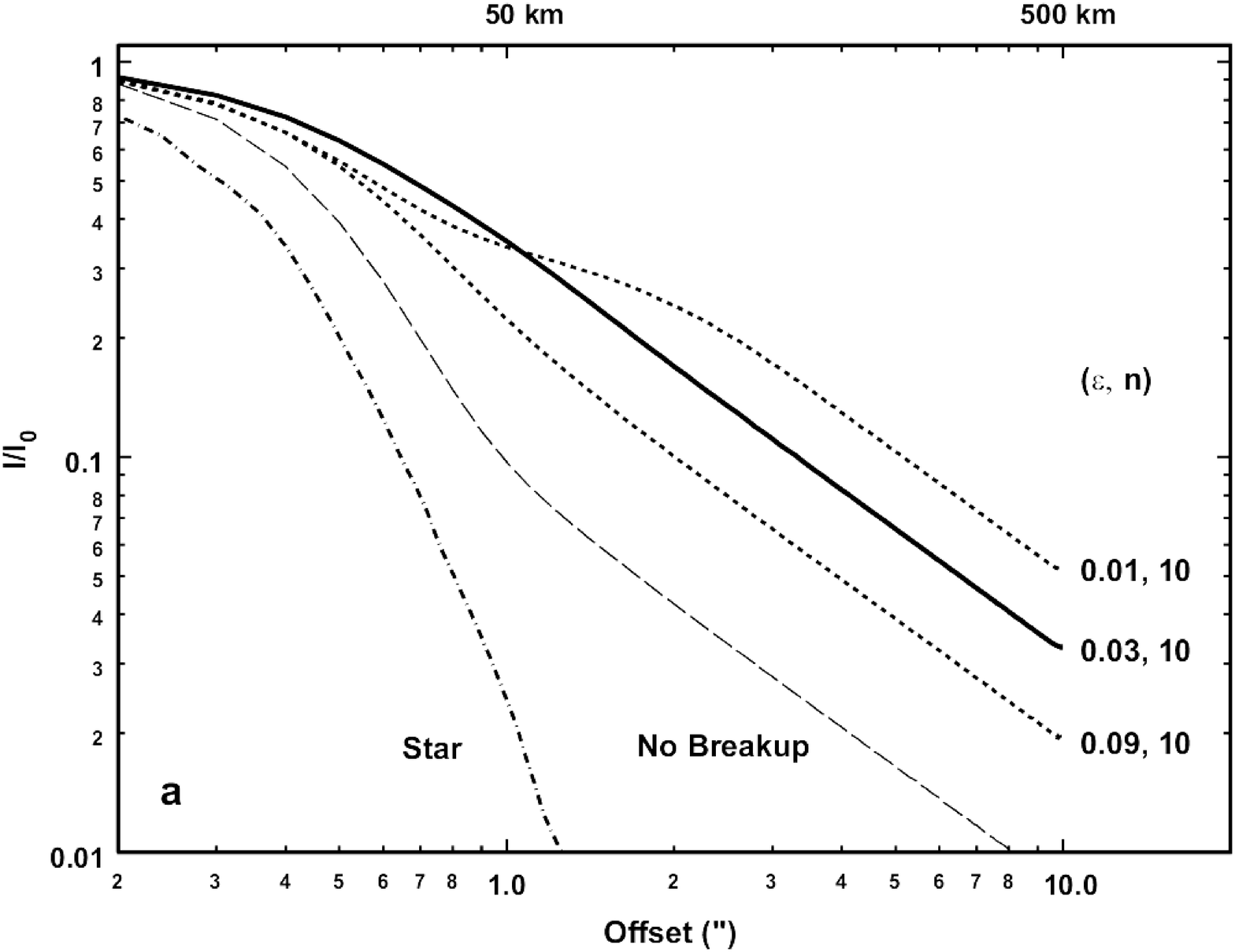}
\plotone{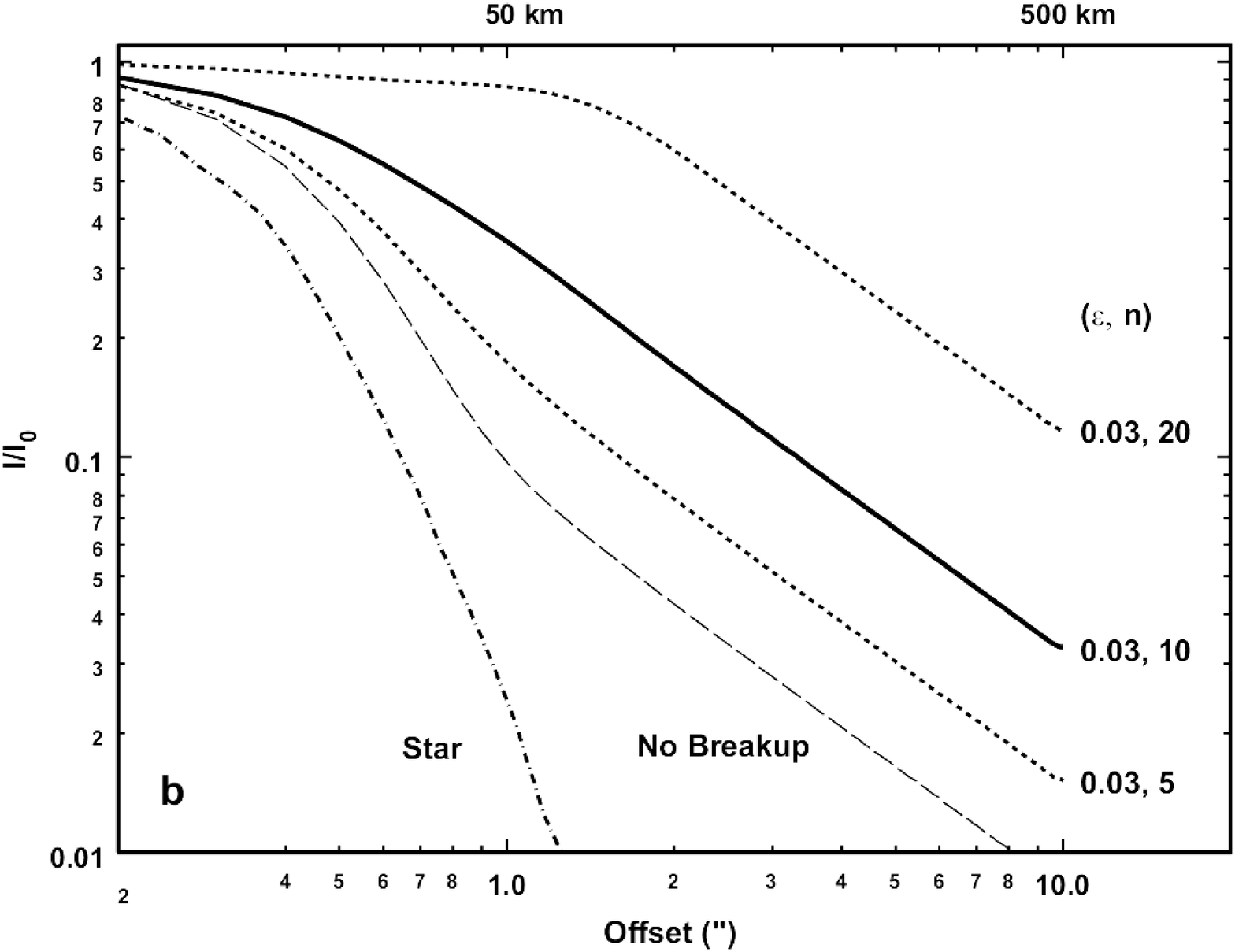}
\caption{Behavior of the grain aggregate breakup model with variations in the two input parameters $\epsilon$ and $n$. (a) Top Panel - Parameter $\epsilon$ is varied and $n$ is kept constant. (b) Bottom Panel - Parameter $n$ is varied and $\epsilon$ is kept constant. The model profiles are compared to the seeing profile of a star and the model results for the case with no breakup ($n=0$). The model with $\epsilon=0.03$ and $n=10$ most closely resembles the observed profiles and is plotted as a solid line.}
\end{figure}

\begin{figure}
\epsscale{0.45}
\plotone{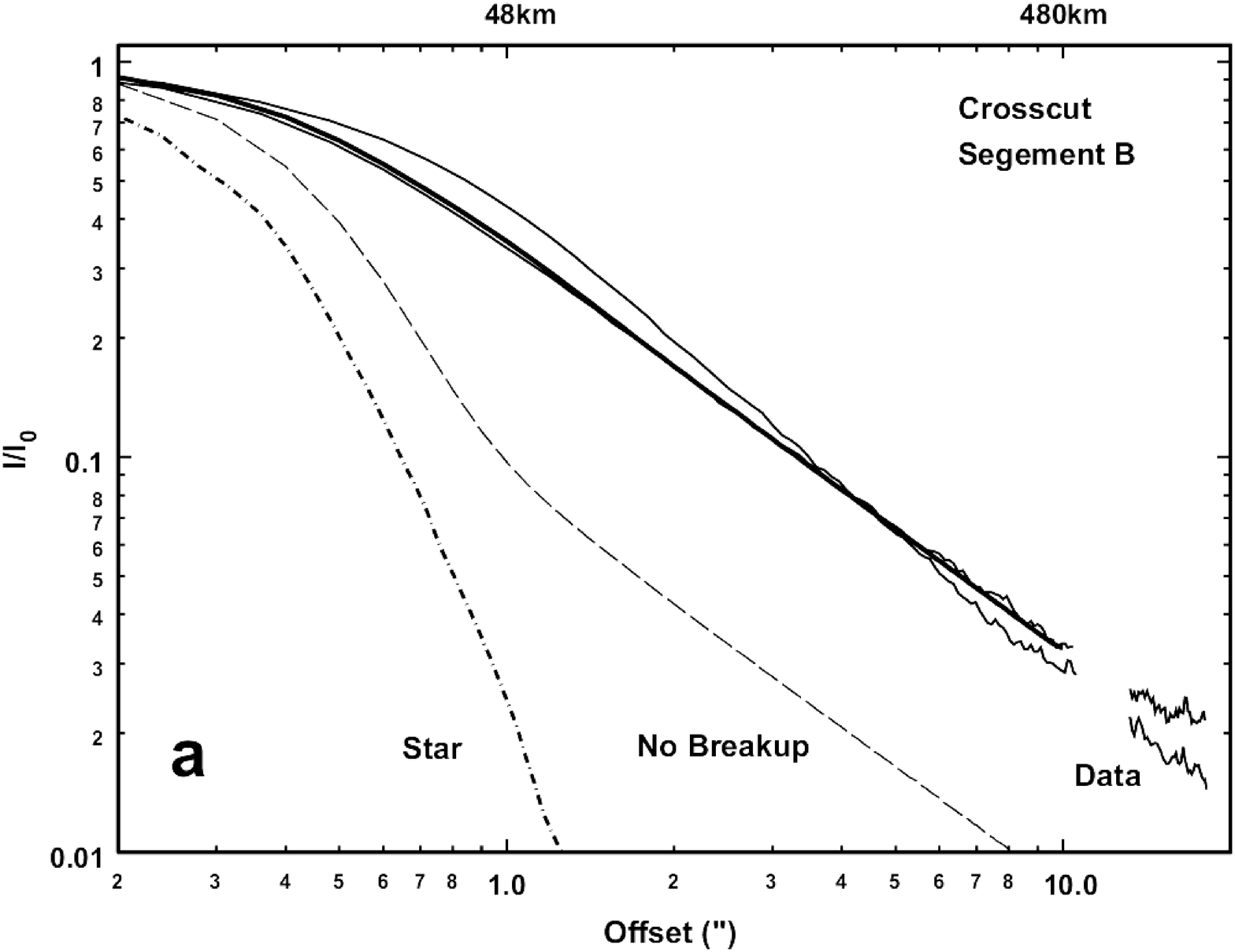}
\plotone{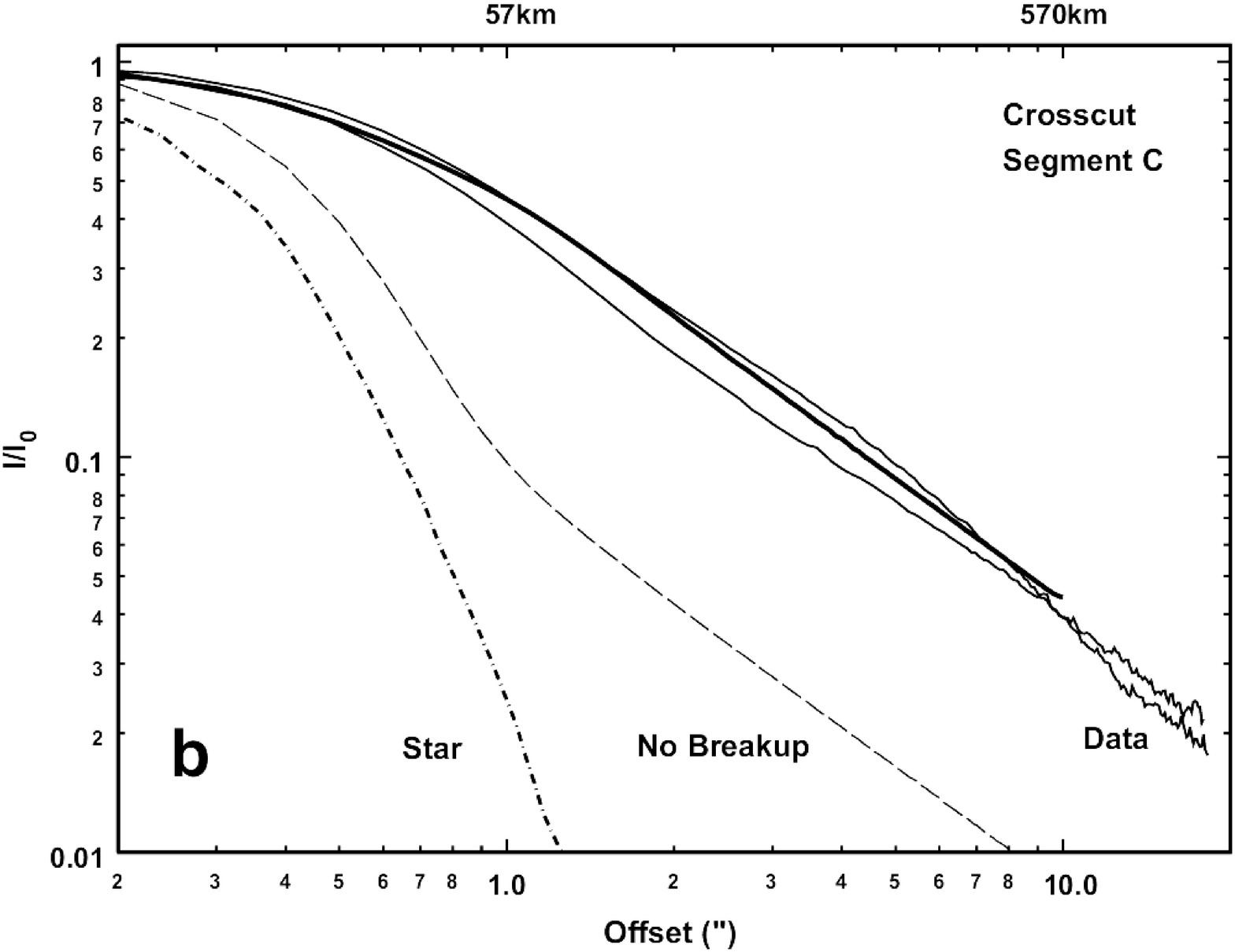}
\plotone{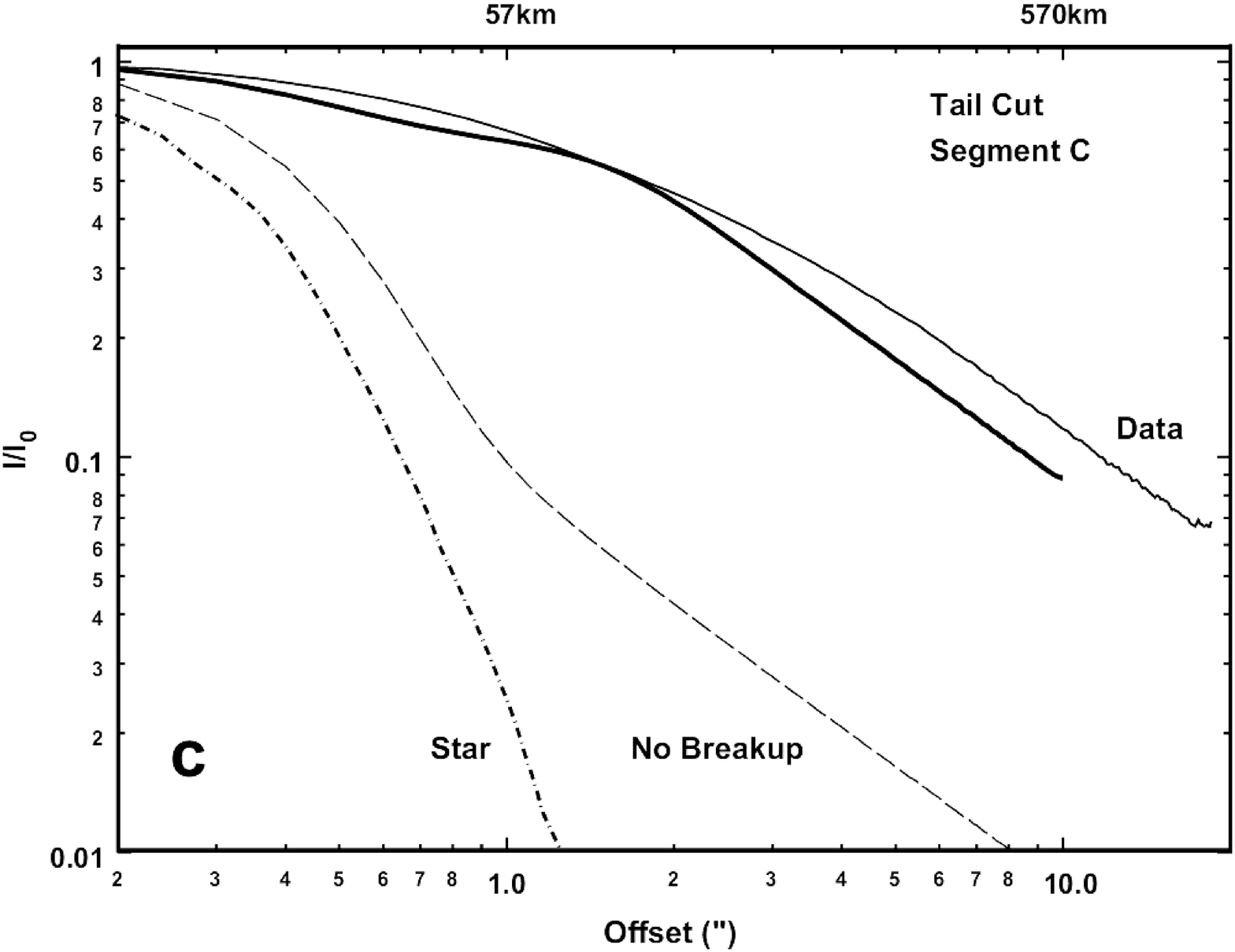}
\caption{Plot of observed and model surface brightness profiles for Comet SW3 in 2006 May in $H$ band. The model fit is shown with the thicker solid line. Panel (a) shows the observed radial profile for the NE component of Segment B compared to a model with $\epsilon = 0.03$ and $n=10$. Panel (b) shows the same for Segment C, except the model parameters are $\epsilon = 0.03$ and $n=12$. Panel (c) shows the radial profile along the tail of Segment C compared to a model with $\epsilon = 0.02$ and $n = 15$.}
\end{figure}

\end{document}